\documentclass[symmetry,article,accept,moreauthors,pdftex]{Definitions/mdpi}

\firstpage{1} 
\pubvolume{14}
\issuenum{2}
\articlenumber{285}
\pubyear{2022}
\copyrightyear{2022}
 \externaleditor{{Academic Editor:  Maxim Y. Khlopov and Ignatios Antoniadis} } 

\datereceived{23 December 2021} 
\dateaccepted{27 January 2022} 
\datepublished{31 January 2022} 
\hreflink{https://\linebreak doi.org/} 

\usepackage[english]{babel}

\usepackage{xcolor}

\usepackage{enumerate}

\usepackage{bm}

\usepackage{graphicx}

\def\ba{\begin{array}}
\def\ea{\end{array}} 
\def\bea{\begin{eqnarray}}
\def\eea{\end{eqnarray}}
\def\beq{\begin{equation}}
\def\eeq{\end{equation}}
\def\ben{\begin{enumerate}}
\def\een{\end{enumerate}}

\def\brr{\begin{array}}
\def\err{\end{array}}

\def\chiC{\chi_\S}

\Title{A Peek Outside Our Universe}
\TitleCitation{A Peek Outside Our Universe}

\Author{Enrique Gaztanaga * $^{1,2}$\orcidA{} and Pablo Fosalba  $^{1,2}$ \orcidB{}}   

\AuthorNames{Enrique Gaztanaga, Pablo Fosalba}

\AuthorCitation{Gaztanaga, E.; Fosalba, P.}

\address{%
$^1$ \quad {Institute} of Space Sciences (ICE, CSIC), 08193 Barcelona, Spain\\ 
$^2$ \quad  {Institut} d Estudis Espacials de Catalunya (IEEC), 08034 Barcelona,  Spain}

\corres{Correspondence: {gaztanaga@darkcosmos.com}}  

\abstract{According to general relativity (GR), a universe with a  cosmological constant $\Lambda$, like ours, is trapped inside an event horizon,  $r< \sqrt{3/\Lambda}$.
What is outside?  We show, using Israel (1967) junction conditions,  that there could be a different universe outside. Our universe looks like a black hole for an outside observer. Outgoing radial null geodesics cannot escape  our universe, 
but  incoming photons can enter and leave an imprint on our CMB sky. 
We present a picture of such a fossil record from the analysis of CMB maps that agrees with the black hole universe predictions but challenges our understanding of the origin of the primordial universe.}


\begin{document}

\section{Introduction}
\label{S:1}


An event horizon (EH) $r_*$, or trapped surface,  for a given observer can be defined as the distance beyond which this observer will never see: $r<r_*$. The most famous EH is, of course, that of a black hole (BH) of mass M. Dimensional analysis tells us that
a relativistic ($c$) gravitational ($G$) system of mass $M$ has an associated EH:
$r_* \simeq GM/c^2$. The same dimensional analysis indicates 
that a cosmological constant $\Lambda$ is associated
with a relativistic EH, $r_* \simeq 1/\sqrt{\Lambda c^2}$, which, in principle, 
is independent of $G$. We will show first how these EHs appear as solutions to Einstein's field equations and next how the different EHs are related in our universe expansion.


The most general form for a metric with spherical symmetry in proper coordinates $x^\mu=(t,r,\delta,\theta)$ with $c=1$ can be written as \cite{Padmanabhan}:
 \beq
 ds^2 =g_{\mu\nu} dx^\mu dx^\nu=  -A(t,r) dt^2 + B(t,r) dr^2 + r^2 d\Omega^2
\label{eq:newFRW}
 \eeq 
 where we have introduced:  $d\Omega^2 = \cos^2\delta d\theta^2 + d\delta^2$.
 Einstein's field equations for that metric in empty space $\rho=p=\Lambda=0$ 
result in the Schwarzschild (SW) metric:
 \beq
 ds^2  = -[1-\frac{2GM}{r} 
 ] ~dt^2 + \frac{dr^2}{1-\frac{2GM}{r} } + r^2 d\Omega^2
\label{eq:Schwarzschild}
 \eeq 
 where $M$ can be interpreted as a singular point mass at $r=0$. 
 As it is well known, the EH at $r_*= 2GM$ prevents us from seeing such a 
 naked singularity \cite{Penrose}. 
 Outgoing null radial geodesics cannot leave the interior of $r_*$, whereas incoming null radial geodesics can cross inside $r_*$ in proper time, even when, for an observer outside, this takes $t=\infty$ in her time.

\subsection{deSitter Metric} 

The solution for Equation~(\ref{eq:newFRW}) with $\rho=p=M=0$ but $\Lambda \ne 0$ is the deSitter (dS) metric:
  \beq
 ds^2= - [1- r^2 H_\Lambda^2 ]~  dt^2 + 
\frac{dr^2}{1- r^2 H_\Lambda^2} + r^2 d\Omega^2
\label{eq:deSitter}
\eeq
which is also static and has 
{an EH, or trapped surface,} at $r_*=1/H_\Lambda$, where
$ H_\Lambda^2 \equiv 8\pi G\rho_\Lambda/3$ and
$ \rho_\Lambda = \Lambda/(8\pi G) + \rho_{vac}$. We include a constant $\rho_{vac}=V=V(\phi)=-p_{vac}$ to account for vacuum energy (or the potential of a trapped scalar field), which is {physically} degenerate with $\Lambda$. 
The inside of $r_*=1/H_\Lambda$ is causally disconnected and corresponds to an EH: radial null events ($ds^2=0$) connecting $(0,r_0)$ with $(t,r)$ 
take $t=\infty$ to reach $r=r_*$ from any point~inside.

\subsection{The FLRW Metric and $\Lambda$}

The Friedmann--Lemaitre--Robertson--Walker (FLRW) flat ($k=0$) metric in spherical comoving coordinates $(\tau,\chi,\delta,\theta)$, corresponds to a homogeneous and isotropic space-time:
\beq
ds^2 = -d\tau^2 + a(\tau)^2\left[ d\chi^2 + \chi^2 d\Omega^2 \right]
\label{eq:frw}
\eeq
where the scale factor $a(\tau)$ describes the expansion/contraction as a function of time.  
For a comoving observer and a perfect fluid,
the field equations reduce to:   
\beq
3H^2 \equiv 3\left(\frac{\dot{a}}{a}\right)^2 =  8\pi G (\rho_m a^{-3}+\rho_R a^{-4}  +  \rho_\Lambda)
\label{eq:Hubble}
\eeq
where  $\rho_m$ and $\rho_R$ are the matter and radiation density 
today ($a=1$) and  $\rho_\Lambda=-p_\Lambda$ is the effective cosmological constant density introduced in the dS metric. Given $\rho$ and $p$ at some time, we can find $a=a(\tau)$ and determine the metric in Equation~(\ref{eq:frw}).
Observations show that the expansion rate today is dominated by $\rho_\Lambda$.
This indicates that the FLRW metric describes the interior of a trapped surface of size $r_*=1/H_\Lambda$, like the dS metric. In fact, both metrics are equivalent in that regime \cite{Mitra2012,hal-03344159}. They also reproduce the steady-state cosmological principle \cite{SteadyState}.

Interpreting $\Lambda$ as a boundary term to the GR equations and imposing the so-called zero action principle, $r_*=1/H_\Lambda$ can be given in terms of ordinary contributions to the energy density, $ \rho_\Lambda= <\rho_m/2+\rho_R>$ 
\cite{Gaztanaga2021,Gaztanaga2020}.

\subsection{SW--FLRW Perturbation}

 The Schwarzschild (SW) metric is commonly used to describe the outside of BHs or stars and it should be understood as a perturbation inside a larger background. 
 An example of this is the Schwarzschild--deSitter (SW--dS) {metric:} 
   \beq
 ds^2 =   - [1 -\frac{2GM}{r} - r^2 H_\Lambda^2 ]~  dt^2 + 
\frac{dr^2}{1 -\frac{2GM}{r} - r^2 H_\Lambda^2} + r^2 d\Omega^2
\label{eq:deSitter2}
\eeq
which corresponds to an exact solution with both {non-zero} $\Lambda$ and $M$.
More generally, the FLRW metric is the background to the SW solution  \cite{Kaloper2010},  e.g.,  it replaces $r=a(\tau) \chi$ in Equation~(\ref{eq:Schwarzschild}).
For large $r$, we recover the dS metric, as in Equation~(\ref{eq:deSitter2}) above, which is  equivalent to the FLRW metric dominated by $\Lambda$ \cite{Mitra2012}. Close to the BH at $r_*<r<1/H_\Lambda$, we recover the SW metric.

\section{Outside Our FLRW Universe}

In proper coordinates $r=a\chi$, 
the FLRW metric
with $H=H(\tau)$ is also trapped inside the same EH as the dS metric, $r_*=1/H_\Lambda$, because $H(\tau)>H_\Lambda$. We can see this by considering outgoing radial null geodesics in the FLRW metric (Equation~(\ref{eq:frw})):
\beq
r_{out} = a(\tau) \int_\tau^\infty \frac{d\tau'}{a(\tau')} =
a \int_a^\infty \frac{d\ln{a'}}{a' H(a')} <\frac{1}{H_\Lambda} = r_*
\label{eq:chi}
\eeq
which shows that signals cannot escape from the inside to the outside of the EH. However, incoming radial null geodesics $a(\tau) \int_0^\tau \frac{d\tau}{a(\tau)} $ could be larger than $r_*$ if we look back in time long enough. This shows that observers living in the interior are trapped inside the EH, but they can, in principle, observe what happened outside.

What is outside $r_*=1/H_\Lambda$ 
in the FLRW metric?  
The FLRW comoving coordinates $(\tau,\chi)$ can be matched to the SW proper coordinates $(t,r)$. The joint metric
is what \cite{Gaztanaga2021,hal-03344159} call a BHU solution. The particular case where the inside is dS (and the outside SW) is called BH.fv (where fv stands for false vacuum). The  BHU metric is also a solution to Einstein's field equations. To prove this, we simply need to 
show that the junction follows the Israel matching conditions \cite{Israel}. 
The two metrics can be matched on a timelike hypersurface $\Sigma$
of constant $\chi$:
\beq
ds^2_{\Sigma}= h_{ab} dy^a dy^b= -d\tau^2 + a^2(\tau) \chiC^2 d\Omega^2
\label{eq:Sigma}
\eeq
and the extrinsic curvature $K$ at $\Sigma$ is the same in both sides.
 The matching conditions $h_{-}=h_{+}$ and $K_{+}=K_{-}$ reduce to \cite{hal-03344159}
\beq
r = R(\tau) =  a(\tau) \chiC
~~ ;~~ 
\dot{R}^2 = R^2 H^2 = \frac{R_*}{R} 
\label{eq:match}
\eeq
where $R_* \equiv 2GM$.
Staring from small $a$, as we increase $\tau$, both $R$ and 
$\dot{R} = H R$ grow until we reach $H R=c= 1$, which corresponds to the event horizon $R_*=2GM =1/H_\Lambda$. It takes $t=\infty$ in  SW time to  asymptotically reach $R_*$. 
This proves that the joint BHU  metric  is also a solution to Einstein's field equations with no surface terms in the junction.  This is equivalent to stating that
the $\Lambda$ term corresponds to a trapped surface $R_*=1/H_\Lambda$ in the FLRW metric which  matches the EH of a BH.
Generalization to $k\ne 1$ is straightforward (see  \S12.5.1 in  \cite{Padmanabhan,Stuckey,hal-03344159}). 

Recall that the SW metric is a perturbation of a larger FLRW metric, i.e, a BH-like metric embedded in a background described by the FLRW  metric. This means that outside  $r_*=1/H_\Lambda$, we have another FLRW metric, like in a Matryoshka doll. From the outside, the inner FLRW metric looks like a BH. There could be many other BHUs  inside and outside  $r_*=1/H_\Lambda$, so the structure could be better described by a fractal.

 \subsection{Causal Structure}

\begin{figure}[H]
\includegraphics[width=0.7\linewidth]{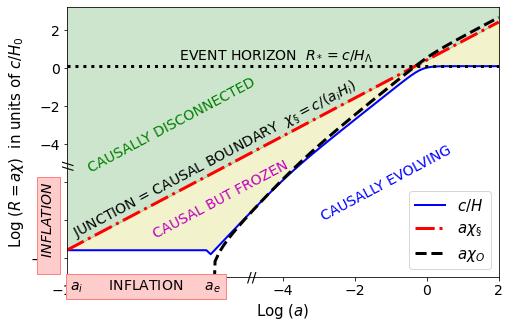}
\caption{
Proper coordinate: $R=a(\tau) \chi$
in units of $c/H_0$ as a function of cosmic time $a$. The Hubble horizon  $c/H $ (blue continuous line) is compared to the observable universe $a\chi_O$ after inflation (dashed line) and the   primordial causal boundary $\chiC = c/(a_i H_i)$ (dot-dashed red line).
Larger scales (green shading) are causally disconnected and smaller scales (yellow shading) are dynamically frozen. 
After inflation, $c/H$  grows again.  At $a \simeq 1$, the Hubble horizon reaches our event horizon $R_* = c /H_\Lambda$. 
At the CMB last scattering, we can observe both frozen and causally disconnected perturbations.}
\label{fig:Vacuum}
\end{figure}

In the FLRW universe, the Hubble horizon $r_H$ is defined as $r_H = c/H$. 
Scales larger than $r_H$ cannot evolve because the time a perturbation takes to travel that distance is larger than the expansion time. 
This means that $r>r_H$ scales are "frozen out" (structure cannot evolve) and are causally disconnected from the rest. 
Thus, $c/H$ represents a dynamical causal horizon that is evolving 
(blue line in Figure~\ref{fig:Vacuum}).

We can sketch the evolution of our universe in Figure~\ref{fig:Vacuum}. A primordial field $\psi$ settles or fluctuates into a false (or  slow-rolling) vacuum which will create a BH.fv  with a junction $\Sigma$ in Equation~(\ref{eq:Sigma}), where the causal boundary is fixed in comoving coordinates and corresponds to the particle horizon during inflation, $\chiC = c/(a_i H_i)$, or the Hubble horizon when inflation begins.
The size $R= a(\tau) \chiC$ of this vacuum grows and asymptotically tends to $R_*=c/H$  following Equation~(\ref{eq:match}) with $H=H_i$. 
The inside of this BH will be expanding exponentially, $a= e^{\tau H_i}$, while the Hubble horizon is fixed at $1/H_i$. 
When this inflation ends \cite{Starobinski1979,Guth1981,Linde1982,Albrecht1982}, 
vacuum energy excess converts into matter and radiation (reheating). This results in BHU, where the infinitesimal Hubble horizon 
starts to grow following the standard Big Bang evolution.
The observable universe (or particle horizon) after inflation, $\chi_O$, is:
\beq
\chi_O = \chi_O(a)=  \int_{a_e}^{a}  \frac{ d\ln a'}{a' H(a')} = \chi_O(1) - \bar{\chi}(a),
\label{eq:chia}
\eeq
where $a_e$ is the scale factor when inflation
ends. For $\Omega_\Lambda \simeq 0.7$, the particle horizon today is
$\chi_O(1) \simeq 3.26 c/H_0$ and $\bar{\chi}(a) = \int_a^1 d\ln a/(aH)$ is the radial lookback time, {which, for a  flat universe, agrees
with the comoving angular diameter distance, $d_A=\bar{\chi}$}. 
{The observable universe becomes larger than $R_*$ when $a > 1$}, as shown in Figure~\ref{fig:Vacuum}
(compare dotted and dashed lines).
{This shows that observers like us, living in the interior of the BH universe, are trapped inside $R_*$ but can nevertheless observe what happened outside}.
We can estimate $\chiC$ from
$\rho_\Lambda = < \rho_m/2 + \rho_R>$ ({see \S I.A)}, where the average is in the lightcone inside $\chiC$. For $\Omega_\Lambda \simeq 0.7$, 
\cite{Gaztanaga2021} found that $\chiC \simeq  3.34 c/H_0$, which is close to $\chi_O$ today. Imagine that $\Omega_\Lambda$ is caused by some dark energy and has nothing to do with  $\chiC$. We still have that $\chiC \lesssim  \chi_O$, because otherwise $\chiC$ would have crossed $RH=1$ early on, 
resulting in a smaller $\chi_O$ than measured (see Figure~\ref{fig:Vacuum}).

Thus, at the time of the CMB last scattering (when $d_A \simeq \chi_O$), $\chiC$ corresponds to an angle $\theta = \chiC/d_A  \lesssim  1 \ $rad$ \ \simeq 60$ deg. Therefore, we can actually observe scales larger than $\chiC$, scales that are not causally connected!
This could be related to the so-called CMB anomalies (i.e, apparent deviations with respect to simple predictions from $\Lambda$CDM, see \cite{P18isotropy,2016CQGra..33r4001S} and references therein) or the parameter tensions in measurements from vastly different cosmic scales or times \cite{P18cosmo,Riess19,des2018,DiValentino}).

\subsection{A Peek Outside}

\begin{figure}[H]
\includegraphics[width=1.0\linewidth]{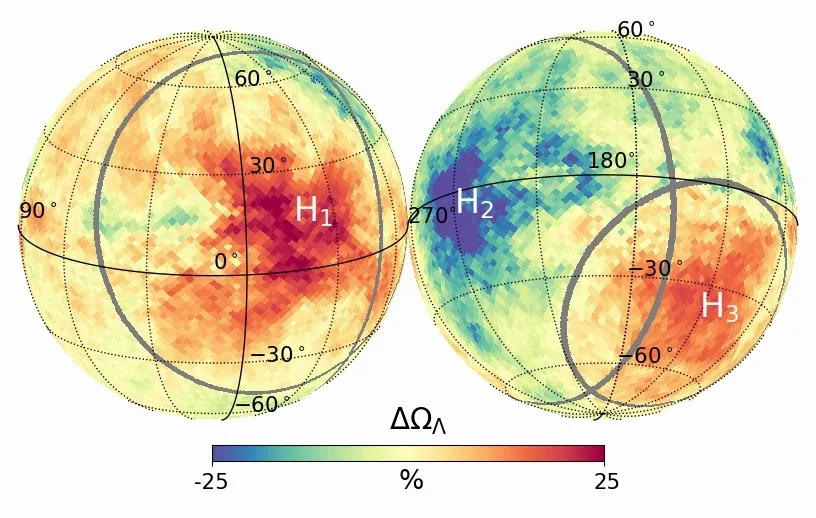}
\caption{Map of the best-fit values of the dark energy density $\Omega_\Lambda$ across the celestial sphere, estimated from partial sky (discs) measurements of the Planck CMB maps. The large grey circles delimit areas across the sky with significantly different values of $\Omega_\Lambda$.}
\label{fig:CMB}
\end{figure}

A recent analysis of the Planck temperature anisotropy data, {Ref}.~\cite{FG20}, 
 shows that the distribution of best-fit dark energy density $\Omega_\Lambda$ exhibits three distinct regions across the CMB sky (marked by the three large grey circles in Figure~\ref{fig:CMB}). 
These regions have radii ranging from 40 to 70 degrees. The sizes of these structures are in agreement with the scale of the causal boundary $\chiC$ for $\Lambda$ dominated universes.
As shown in \cite{FG20}, the size of each of these regions is correlated with the mean value of $\Omega_\Lambda$ over that portion of the sky, in good agreement with the BHU prediction.
The same large-scale anisotropic patterns are observed for the distribution of other basic $\Lambda$CDM  cosmological parameters.
This represents a very significant break-down of the main hypothesis of the Big Bang model: the assumption that the universe is isotropic on a large scale. 
The observed anisotropy has a tiny probability (of order $10^{-9}$) of being a Gaussian statistical fluctuation of an otherwise isotropic universe~\cite{FG20}.

In summary, Figure~\ref{fig:CMB} shows that:
(a) regions with a given value of $\Omega_\Lambda$ have a corresponding angular size $\theta$ that agrees with the BHU prediction (see Figure~31 in \cite{FG20}),
(b)~causally disconnected regions of the sky (larger than $\chiC$)
fit the same physical model of acoustic oscillations very well, and
(c) the background is similar but with parameters that are significantly different across disconnected regions.
This suggests that the underlying physical mechanism sourcing the observed  anisotropy encompasses scales beyond our causal universe. This is in apparent tension with simple models of inflation (as sources of perturbation for the largest scales) and
opens the door to revisiting our basic understanding of the origin of the primordial universe \cite{universe}.

\vspace{-6pt}

\acknowledgments{
{This work has been }supported by Spanish MINECO  grants PGC2018-102021 and ESP2017-89838-C3-1-R and EU grants LACEGAL 734374 and EWC 776247 with ERDF funds and grant 2017-SGR-885 of the Generalitat de Catalunya.} 

\conflictsofinterest{The authors declare no conflict of interest.}


\begin{adjustwidth}{-\extralength}{0cm}
\reftitle{References}


\begin{thebibliography}{999}

\bibitem[{Padmanabhan}(2010)]{Padmanabhan}
{Padmanabhan}, T.
\newblock {\em {{Gravitation: foundations and frontiers}}};   {Cambridge University Press}: Cambridge, 2010.   


\bibitem[{Penrose}(1969)]{Penrose}
{Penrose}, R.
\newblock {Gravitational Collapse: The Role of General Relativity}.
\newblock {\em Nuovo Cimento Rivista Serie} {\bf 1969}, {\em 1},~252.

\bibitem[{Mitra}(2012)]{Mitra2012}
{Mitra}, A.
\newblock {Interpretational conflicts between the static and non-static forms
  of de Sitter metric}.
\newblock {\em Sci. Rep.} {\bf 2012}, {\em 2},~923.
\newblock
  doi:{\changeurlcolor{black}\href{https://doi.org/10.1038/srep00923}{\detokenize{10.1038/srep00923}}}.

\bibitem[Gaztanaga(2021)]{hal-03344159}
Gaztanaga, E.d.
\newblock {The Black Hole Universe (BHU) from a FLRW Cloud}.
\newblock Submitted to Physics of the Dark Universe. 
 Available online: \url{https://hal.archives-ouvertes.fr/hal-03344159} 
  accessed on September 14, 2021.


\bibitem[{O'Raifeartaigh} and {Mitton}(2015)]{SteadyState}
{O'Raifeartaigh}, C.; {Mitton}, S.
\newblock {A new perspective on steady-state cosmology}. \emph{arXiv} {\bf 2015},
 arXiv:1506.01651.

\bibitem[{Gazta{\~n}aga}(2021)]{Gaztanaga2021}
{Gazta{\~n}aga}, E.
\newblock {The cosmological constant as a zero action boundary}.
\newblock {\em Mon. Not. R. Astron. Soc.} {\bf 2021}, {\em 502},~436--444.
\newblock
  doi:{\changeurlcolor{black}\href{https://doi.org/10.1093/mnras/stab056}{\detokenize{10.1093/mnras/stab056}}}.

\bibitem[{Gaztanaga}(2020)]{Gaztanaga2020}
{Gaztanaga}, E.
\newblock {The size of our causal Universe}.
\newblock {\em Mon. Not. R. Astron. Soc.} {\bf 2020}, {\em 494},~2766--2772.
\newblock
  doi:{\changeurlcolor{black}\href{https://doi.org/10.1093/mnras/staa1000}{\detokenize{10.1093/mnras/staa1000}}}.

\bibitem[{Kaloper} \em{et~al.}(2010){Kaloper}, {Kleban}, and
  {Martin}]{Kaloper2010}
{Kaloper}, N.; {Kleban}, M.; {Martin}, D.
\newblock {McVittie's legacy: Black holes in an expanding universe}.
\newblock {\em Phys. Rev. D} {\bf 2010}, {\em 81},~104044.
\newblock
  doi:{\changeurlcolor{black}\href{https://doi.org/10.1103/PhysRevD.81.104044}{\detokenize{10.1103/PhysRevD.81.104044}}}.

\bibitem[{Israel}(1967)]{Israel}
{Israel}, W.
\newblock {Singular hypersurfaces and thin shells in general relativity}.
\newblock {\em Nuovo Cimento B Serie} {\bf 1967}, {\em 48},~463--463.
\newblock
  doi:{\changeurlcolor{black}\href{https://doi.org/10.1007/BF02712210}{\detokenize{10.1007/BF02712210}}}.

\bibitem[{Stuckey}(1994)]{Stuckey}
{Stuckey}, W.M.
\newblock {The observable universe inside a black hole}.
\newblock {\em Am. J. Phys.} {\bf 1994}, {\em 62},~788--795.
\newblock
  doi:{\changeurlcolor{black}\href{https://doi.org/10.1119/1.17460}{\detokenize{10.1119/1.17460}}}.

\bibitem[{Starobinski{\v{i}}}(1979)]{Starobinski1979}
{Starobinski{\v{i}}}, A.A.
\newblock {Spectrum of relict gravitational radiation and the early state of
  the universe}.
\newblock {\em Sov. Exp. Theor. Phys. Lett.} {\bf 1979}, {\em
  30},~682.

\bibitem[{Guth}(1981)]{Guth1981}
{Guth}, A.H.
\newblock {Inflationary universe: A possible solution to the horizon and
  flatness problems}.
\newblock {\em Phys. Rev. D} {\bf 1981}, {\em 23},~347--356.
\newblock
  doi:{\changeurlcolor{black}\href{https://doi.org/10.1103/PhysRevD.23.347}{\detokenize{10.1103/PhysRevD.23.347}}}.

\bibitem[{Linde}(1982)]{Linde1982}
{Linde}, A.D.
\newblock {A new inflationary universe scenario: A possible solution of the
  horizon, flatness, homogeneity, isotropy and primordial monopole problems}.
\newblock {\em Phys. Lett. B} {\bf 1982}, {\em 108},~389--393.
\newblock
  doi:{\changeurlcolor{black}\href{https://doi.org/10.1016/0370-2693(82)91219-9}{\detokenize{10.1016/0370-2693(82)91219-9}}}.

\bibitem[{Albrecht} and {Steinhardt}(1982)]{Albrecht1982}
{Albrecht}, A.; {Steinhardt}, P.J.
\newblock {Cosmology for Grand Unified Theories with Radiatively Induced
  Symmetry Breaking}.
\newblock {\em Phys. Rev. Lett.} {\bf 1982}, {\em 48},~1220--1223.
\newblock
  doi:{\changeurlcolor{black}\href{https://doi.org/10.1103/PhysRevLett.48.1220}{\detokenize{10.1103/PhysRevLett.48.1220}}}.

\bibitem[{Planck Collaboration}(2020)]{P18isotropy}
{Planck Collaboration}.
\newblock {Planck 2018 results. VII. Isotropy and statistics of the CMB}.
\newblock {\em Astron. Astrophys.} {\bf 2020}, {\em 641},~A7.
\newblock
  doi:{\changeurlcolor{black}\href{https://doi.org/10.1051/0004-6361/201935201}{\detokenize{10.1051/0004-6361/201935201}}}.

\bibitem[{Schwarz} \em{et~al.}(2016){Schwarz}, {Copi}, {Huterer}, and
  {Starkman}]{2016CQGra..33r4001S}
{Schwarz}, D.J.; {Copi}, C.J.; {Huterer}, D.; {Starkman}, G.D.
\newblock {CMB anomalies after Planck}.
\newblock {\em Class. Quantum Gravity} {\bf 2016}, {\em 33},~184001.
\newblock
  doi:{\changeurlcolor{black}\href{https://doi.org/10.1088/0264-9381/33/18/184001}{\detokenize{10.1088/0264-9381/33/18/184001}}}.

\bibitem[{Planck Collaboration}(2020)]{P18cosmo}
{Planck Collaboration}.
\newblock {Planck 2018 results. VI. Cosmological parameters}.
\newblock {\em Astron. Astrophys.} {\bf 2020}, {\em 641},~A6.
\newblock
  doi:{\changeurlcolor{black}\href{https://doi.org/10.1051/0004-6361/201833910}{\detokenize{10.1051/0004-6361/201833910}}}.

\bibitem[{Riess}(2019)]{Riess19}
{Riess}, A.G.
\newblock {The expansion of the Universe is faster than expected}.
\newblock {\em Nat. Rev. Phys.} {\bf 2019}, {\em 2},~10--12.
\newblock
  doi:{\changeurlcolor{black}\href{https://doi.org/10.1038/s42254-019-0137-0}{\detokenize{10.1038/s42254-019-0137-0}}}.

\bibitem[{DES Collaboration}(2019)]{des2018}
{DES Collaboration}.
\newblock {Cosmological Constraints from Multiple Probes in the Dark Energy
  Survey}.
\newblock {\em Phys. Rev. Lett.} {\bf 2019}, {\em 122},~171301. 
  doi:{\changeurlcolor{black}\href{https://doi.org/10.1103/PhysRevLett.122.171301}{\detokenize{10.1103/PhysRevLett.122.171301}}}.

\bibitem[{Di Valentino} \em{et~al.}(2021){Di Valentino}, {Mena}, {Pan},
  {Visinelli}, {Yang}, {Melchiorri}, {Mota}, {Riess}, and {Silk}]{DiValentino}
{Di Valentino}, E.; {Mena}, O.; {Pan}, S.; {Visinelli}, L.; {Yang}, W.;
  {Melchiorri}, A.; {Mota}, D.F.; {Riess}, A.G.; {Silk}, J.
\newblock {In the realm of the Hubble tension-a review of solutions}.
\newblock {\em Class. Quantum Gravity} {\bf 2021}, {\em 38},~153001.
\newblock
  doi:{\changeurlcolor{black}\href{https://doi.org/10.1088/1361-6382/ac086d}{\detokenize{10.1088/1361-6382/ac086d}}}.

\bibitem[{Fosalba} and {Gazta{\~n}aga}(2021)]{FG20}
{Fosalba}, P.; {Gazta{\~n}aga}, E.
\newblock {Explaining cosmological anisotropy: Evidence for causal horizons
  from CMB data}.
\newblock {\em Mon. Not. R. Astron. Soc.} {\bf 2021}, {\em 504},~5840--5862.
\newblock
  doi:{\changeurlcolor{black}\href{https://doi.org/10.1093/mnras/stab1193}{\detokenize{10.1093/mnras/stab1193}}}.
 \bibitem[Gaztanaga(2022)]{universe}
Gaztanaga, E.
\newblock {How the Big Bang End Up Inside a Black Hole}.
\newblock {\em Universe} Available as preprint:
\url{https://www.preprints.org/manuscript/202201.0459/v1} 
 accessed on Feb 1, 2022.
 
\end{thebibliography}
\end{adjustwidth}

\end{document}